\documentclass[twocolumns]{mn2e}

\input psfig.sty

\def\T1{\ {$T_1$}\ }
\def\MT1{\ {$M_{T_1}$}\ }
\def\ct1{\ {$(C-T_1)$}\ }
\def\CT10{\ {$(C-T_1)_0$}\ }

\def\VI0{\ {$(V-I)_0$}\ }

\def\2cd{\ {two-color diagram}\ }

\def\sf{\ {specific frequency}\ }
\def\ell{\ {elliptical}\ }

\def\gtsim{\ {\raise-0.5ex\hbox{$\buildrel>\over\sim$}}\ }
\def\ltsim{\ {\raise-0.5ex\hbox{$\buildrel<\over\sim$}}\ }

\begin{document}

\title[The Fornax AMR]{The age-metallicity relationship in the Fornax spheroidal dwarf galaxy}

\author[Piatti et al.]{Andr\'es E. Piatti$^{1,2}$, Andr\'es del Pino$^3$, Antonio Aparicio$^{3,4}$,
\newauthor and Sebasti\'an L. Hidalgo$^{2,3}$ \\
$^1$ Observatorio Astron\'omico, Universidad Nacional de C\'ordoba, Laprida 854, 5000 C\'ordoba, Argentina.\\
$^2$ Consejo Nacional de Investigaciones Cient\'{\i}ficas y T\'ecnicas, Av. Rivadavia 1917, C1033AAJ,
Buenos Aires, Argentina \\
$^3$ Instituto de Astrof\'{\i}sica de Canarias, V\'{\i}a L\'actea s/n, E38200 - La Laguna, Tenerife,
Canary Islands, Spain.\\
$^4$2 University of La Laguna. Avda. Astrof\'{\i}sico Fco. S\'achez, s/n. E38206, La Laguna, Tenerife, Canary 
Islands, Spain.
} 

\maketitle

\begin{abstract}
We produce a comprehensive field star age-metallicity relationship (AMR) from the earliest epoch until $\sim$ 1 Gyr 
ago for three fields in the Fornax dSph galaxy by using $VI$ photometric data obtained 
with $FORS1$ at the $VLT$. We find that the innermost one 
does not contains dominant very old stars (age $>$ 12 Gyr), whereas the relatively outer field does not account for representative
star field populations younger than $\sim$ 3 Gyr. When focusing on the most prominent stellar
populations, we find that the derived AMRs are engraved by the evidence of a outside-in star formation process. 
The studied fields show bimodal metallicity distributions peaked at [Fe/H] = (-0.95 $\pm$ 0.15) dex 
and (-1.15 or -1.25 $\pm$ 0.05) dex, respectively, but only during the first half
of the entire galaxy lifetime. Furthermore, the more metal-rich population appears to be more numerous
in the outer fields, while in the 
innermost Fornax field the contribution of both metallicity populations seems to be similar.
We also find that the metallicity spread $\sim$ 6 Gyr ago is remarkable large, while the intrinsic 
metallicity dispersion at $\sim$ 1-2 Gyr results smaller than that for the relatively older generations of stars. 
We interpret these outcomes as a result of a posssible merger of two galaxies that would have triggered 
a star formation bursting process that peaked between 
$\sim$ 6 and 9 Gyr ago, depending on the position of the field in the galaxy.
\end{abstract}

\begin{keywords}
techniques: photometric -- galaxies: dwarf -- galaxies: individual: Fornax 
\end{keywords}

\section{Introduction}

In the widely accepted $\Lambda$ Cold Dark Matter ($\Lambda$CDM) cosmological scenario, dwarf galaxies
 are key pieces in the 
galaxy formation and evolution puzzle. Their evolution is very likely affected by numerous processes
 such interactions
 with other systems, supernovae feedback, cosmic reionization, among others. Characterizing 
their evolution could 
shed light on the physical mechanisms involved.

The Milky Way (MW) satellites provide a unique opportunity to study in detail galaxy formation and
chemical evolution processes. 
Considered traditionally as relatively simple systems, dwarf spheroidals (dSphs) are so far the most 
common types within the MW
 companions. Nevertheless, the complexity of these galaxies is becoming more evident as new 
detailed and deep data are available.

The Fornax dSph, located at a distance of 136$\pm$5 kpc \cite{mg03,getal07,tetal08,petal08,getal09}, 
is after the Sagittarius dSph the largest and most luminous of the MW companions. These two galaxies are 
the only dSph MW's satellites hosting globular clusters. Particularly, Fornax is up to date known to have 
a very complex 
structure. Two star clumps located at 17$\arcmin$ and at 1.3$\degr$ from the galaxy centre have
 stoked a discussion about whether Fornax suffered a merger. Coleman et al. \shortcite{cetal04},
Coleman \& Da Costa \shortcite{cetal05}, and Coleman \& de Jong \shortcite{cetal08}
proposed that these overdensities are shell structures resulting from a merger with a 
smaller gas-rich system that occurred 2 Gyr ago. The merger scenario is also supported by  
Amorisco \& Evans \shortcite{ae12}  who found  distinct radial velocity stellar components with different 
metallicities, suggesting that 
Fornax is a merger of a bound pair.

On the another hand, de Boer et al. \shortcite{detal13} claimed that these 
clumps are more 
likely the result of the quiet infall of gas 
previously expelled by Fornax during its star formation episodes.
Likewise, previous studies \cite{cetal08,detal12} have assessed important variations in the star formation history
depending on the galactocentric radius. Furthermore, del Pino et al. \shortcite{detal13} found a delay in the main star 
formation burst at the centre of the galaxy, which is compatible with both an outside-in formation and a
merger scenario. As can be figure out, the Age-Metalliciy Relation (AMR) of Fornax should also be engraved by
the different formation events that have taken place during its entire lifetime.

In this paper we revisit the data analysed by del Pino et al. \shortcite{detal13} by making use of a different 
approach \cite{petal12,pg13}. We produced the presently observed AMR for the dominant stellar populations of the
galaxy from its birth until $\sim$ 1 Gyr ago; being able to disentangle a bimodal distribution in the metallicity. 
The latter may be the first direct evidence of a merger formation scenario for the Fornax dSph. In Sect. 2 we
describe the data handling in order to obtain mean ages and metallicities for the representative stellar populations, 
while Sect. 3 deals with the construction of the Fornax's AMR. Sect. 4 focuses on the discussion of the resulting
AMR. Finally, Sect. 5 summaries our main conclusions.

\section{Data handling and scope}

We obtained $VI$ photometric data in three fields of Fornax with $FORS1$ at the $VLT$. We refer the reader 
to del Pino et al. \shortcite{detal13} for details about the observations, reduction, and analysis of the 
data. Here we derive the AMR for the three Fornax fields using the 
procedure applied by Piatti \& Geisler \shortcite{pg13} to produce a comprehensive field star 
AMR of the Large Magellanic Cloud from its earliest epoch until $\sim$ 1 Gyr ago. Briefly, the procedure is 
based on the so-called "representative" population, assumed that the observed Main Sequence (MS) in each field 
is the result of the superposition of MSs with different turnoffs (ages) and 
constant luminosity functions. This "representative" AMR differs from those derived from modeled Star Formation
 Histories (SFHs) in the fact 
that it does not include complete information on all stellar populations, but accounts for the dominant 
population present in each field. Minority populations  are not considered, nor dominant populations younger
than $\sim$ 1 Gyr, due to our inability to age-date them.

Following the precepts outlined by Piatti, Geisler \& Mateluna \shortcite{petal12}, we first subdivided the 
IC1, IC2, and OC fields (see Fig. 1 by del Pino et al. 2013) into
16, 20, and 12 subfields as is shown in Fig. 1.  For each subfield we built the MS luminosity function by 
counting 
the number of stars in $V_o$ bins of 0.25 mag. The chosen 
bin size encompasses the $V_o$ magnitude errors ($\sigma$($V_o$) $\la$ 0.2 mag for $V_o$ $\sim$ 25.9 mag) 
of the stars in each bin, thus producing an appropriate sample of the stars.
%We also assume, for simplicity, the same metallicity for all MS stars within a given subfield. 
Hence, the difference between the number of stars of two adjacent magnitude 
intervals gives the intrinsic number of stars belonging to the faintest interval. Consequently, the maximum
of the distribution function of all these differences in terms of $V_o$  (the differential luminosity 
function) is directly related to 
the most populated TO. Similarly, following the procedure described in Piatti, Geisler \& Mateluna 
\shortcite{petal12}, we measured the $V_o$ magnitudes of the Red Clump (RC) stars, which are relatively
invariant to population effects such as age and metallicity for such stars.
RCs are used in age estimates based on the magnitude difference $\delta$ between the clump/HB and the TO 
for intermediate-age and old clusters. Fig. 2 shows the colour-magnitude diagram for the IC1\_1 subfield with
the purpose of illustrating the method. We have included in the left-hand panel the normalized and
differential MS luminosity functions represented by thin and thick solid lines, and the RC luminosity
distribution by a dotted line as well. Tables 1 and 2 present the derived representative $V_o$(MSTO) and $V_o$(RC) 
magnitudes for the studied Fornax subfields. According to del Pino et al. \shortcite{detal13}, the 
derived $V_o$(MSTO) mags result brighter than the $V_o$ mag at the 90$\%$ completeness level, so that we actually
reach the MSTO of the representative oldest populations of the galaxy. The $V_o$(MSTO) and $V_o$(RC) dispersions
 have been calculated
bearing in mind the broadness of the differential luminosity function and the $V_o$(RC) distribution, 
instead of the 
photometric errors at $V_o$(MSTO) and $V_o$(RC) mags, respectively. The former are clearly larger, 
and represent in general a satisfactory estimate of the spread around the prevailing population, although 
some individual subfields have slightly larger spreads. These larger age spreads should 
not affect the subsequent results.

In order to calculate the representative ages, we first used the values listed in Tables 1 and 2 to  compute
the difference $\delta V_o$ = $V_o$(MSTO) - 
$V_o$(RC),  and then calculated the representative ages and their dispersions by using 
equations (3) and (4) of Geisler et al. \shortcite{getal97} as follows:

\begin{equation}
age(Gyr) = 0.538 + 1.795\delta V_o - 1.480(\delta V_o)^2 + 0.626(\delta V_o)^3
\end{equation}

This equation is only calibrated
for ages larger than 1 Gyr, so that we are not able to produce ages for younger representative populations.
In addition, we also estimated representative metallicities using the equation:

\begin{equation}
[Fe/H] = -15.16 + 17.0(V-I)_{o,-3} - 4.9(V-I)^2_{o,-3}
\end{equation} 

\noindent of Da Costa \& Armandroff \shortcite{dca90}, once the $(V-I)_o$ colours of the
 Red Giant Branch (RGB) at $M_I$ = -3.0 mag and their dispersions were obtained (typically $\sigma$$(V-I)_{o,-3}$ 
= 0.02 mag). The $(V-I)_o$ colours were derived from the intersection of the RGBs 
traced for each subfield and the horizontal line at $M_I$ = -3.0 mag, as illustrated in Fig. 3. 
Note that the adopted subfield size led to obtain well-defined RGBs - without any noticeable
colour (age) spread -, so that the representative metallicities could
be estimated from those RGBs. Table 4 provides with the derived reddening  corrected $(V-I)_o$  values
($E(V-I)$=0.028 mag, Del Pino et al. 2013).

\section{The age-metallicity relationship}

The estimated values of age and metallicity with their respective dispersions for the 48 studied subfields 
in Fornax are treated hereafter as individual point spread functions in order to disentangle the intrinsic AMRs.
In general, one of the unavoidable complications in analysing measured ages and metallicities is that they have 
associated dispersions. 
Indeed, the resulting AMR can differ appreciably depending on whether it is obtained by using only mean values. 
Furthermore, even if dispersions did not play an 
important role, the binning of the age/metallicity
ranges could however bias the results. Thus, for example, by using a fixed age bin size is not appropriate for yielding 
the intrinsic age distribution, since 
the result would depend on the chosen age interval \cite{p10}.
A more robust age bin should be of the order of the age dispersions in that interval.
In the star cluster arena, where the dispersion comes from the uncertainty in measuring ages, this would lead to the selection 
of very narrow bins for young clusters and relatively broader age bins for the older ones. Notice that
the dispersions of the representative ages and metallicities quoted in Tables 3 and 5 refer to the intrinsic
age and metallicity spread among the prevailing stellar populations in the respective subfields.

We then searched Table 3 to find that 
typical age dispersions are  0.10 $\la$ $\Delta$log($t$) $\la$ 0.15. Therefore, we produced the AMR of 
the Fornax field population by setting the age bin sizes according to this logarithmic law, which
traces the variation in the derived age spread in terms of the measured ages. We used 
intervals of $\Delta$log($t$) = 0.10. We proceeded in a similar way when binning
the metallicity range. In this case, we adopted a [Fe/H] interval of 0.25 dex. 
%Thus, the subdivision of the 
%whole age and metallicity ranges was then performed on an observational-based foundation, 
%since the (age,[Fe/H]) dimensions are determined by the typical errors for each age/metallicity range. 
However, there is still an additional issue to be
considered: even though the (age,[Fe/H]) bins are set to match the age/metallicity spread,  
any individual point in the 
AMR plane may fall in the respective (age,[Fe/H]) bin or in any of the eight adjacent bins. This happens
when an (age,[FeH]) point does not fall in the bin centre and, due to its point spread function,  has 
the chance to fall outside it. Note that, since we chose bin dimensions as large as the involved dispersions, 
such points should not fall on average far beyond the adjacent bins. 
We have  taken all these effects into account to produce the AMR of the studied IC1, IC2, and OC Fornax fields
using the procedure proposed by Piatti \& Geisler \shortcite{pg13}. We weighed the contribution of each (age,[Fe/H]) 
pair to each one of the AMR grid bins occupied by it due to their point spread functions. We are confident 
that our analysis yields  accurate morphology and position of the main features in the derived AMRs. 

Fig. 4 shows the resulting presently observed IC1, IC2, and OC AMRs as labelled at the bottom-left
 margin of each panel.
 It is important to keep in mind that each (age,[Fe/H]) point used is 
simply the representative, most dominant population in that subfield, independently of whether it is
primordial or recently formed. These prevailing populations trace
the present-day AMR of the galaxy. They account for the most
important metallicity-enrichment processes that have undergone
in the galaxy lifetime. Minority stellar populations not following
these main chemical galactic processes are discarded. Therefore,
presently-subdominant populations in certain locations could have
been in the majority in the galaxy in the past, but were not considered. 
This could be the case of old stellar populations placed
in the innermost regions. However, unless the Fornax has had an original 
metallicity gradient during its birth, the metallicity of the oldest
stellar populations is recovered from the dominant oldest populations. 
The age error bars in Fig. 4 follow the law $\sigma$log($t$) = 0.10, whereas the [Fe/H]
error bars come from the full width at half-maxima (FWHMs) we derived by fitting Gaussian functions to the metallicity 
distribution in each age interval. The size of the open boxes centred on the mean (age,[Fe/H]) values 
is  a measure of the number of subfields used to compute them. They come from considering the total
number of subfields involved in a mean (age,[Fe/H]) estimate, normalized to the total number of subfield used.
Thus, the larger the number of subfields employed ($\sim$ galaxy mass) to estimate a mean (age,[Fe/H]) value, the 
bigger the size of the box. We also included as a reference a redshift scale in the upper axis, computed assuming
$H_o$ = 70.5 km$\times$s$^{-1}$Mpc$^{-1}$, $\Omega$$_m$ = 0.273, and a flat universe with $\Omega$$_\Lambda$ = 1 -
$\Omega$$_m$. 

The AMRs derived by del Pino et al. \shortcite{detal13} are also shown as background Hess diagrams,
where the most numerous the stellar populations the bluer the contour colour.
The former provides with a complete solution for the SFH along the galaxy lifetime, 
and hence for the AMR. The present method does not give the same information, since most 
of the stellar populations do not appear in the solution but only those that at the present 
time are the representative ones, i.e., those with the presently highest number of stars.

%The fit of a single Gaussian per age bin was performed using the NGAUSSFIT routine in the STSDAS/IRAF\footnote{IRAF is 
%distributed by the National Optical Astronomy Observatories, 
%which is operated by the Association of Universities for Research in Astronomy, Inc., under contract with 
%the National Science Foundation} package. The centre 
%of the Gaussian, its amplitude and its FWHM acted as variables, while the constant and the linear terms were fixed to zero, 
%respectively. We used Gaussian fits for simplicity. We estimated a difference from Gaussian distributions of 
%only $\approx$ 8 $\%$.

\section{Discussion}

Fig. 4 shows that, although at first glance we find stars at any age in the three Fornax fields, the innermost one 
(IC1 field)
does not contains dominant very old stars (age $>$ 12 Gyr) -they could certainly 
still be present as minority population though-, whereas the relatively outer IC2 field does not account for representative
star field populations younger than $\sim$ 3 Gyr. Furthermore, when focusing on the biggest boxes of the three pannels, 
it seems that most of the OC field stars have been formed between 8 and 12 Gyr ago, whereas comparable signifivative
star formation has occurred between 6 and 10 Gyr for IC2, and between 5 and 8 Gyr for IC1. Both results lead us to
conclude that the derived AMRs are engraved by the evidence of a outside-in star formation process in Fornax.

On the metallicity side, Fig. 4 shows bimodal distributions in the three Fornax fields, but only during the first half
of its entire lifetime. In order to draw such a conclusion we required the fulfillment of the following statistical 
criterion : $\sigma$([Fe/H]$_1$) + $\sigma$([Fe/H]$_2$) $>$ $|$[Fe/H]$_1$ -[Fe/H]$_2$$|$), where $\sigma$([Fe/H]$_i$), $i$ = 1,2,
represents the intrinsic spread observed in the mean representative metallicity [Fe/H]$_i$.  
As judged from the mean position of the boxes, such metallicity distributions peak at [Fe/H] =
(-0.95 $\pm$ 0.15) dex and (-1.15 or -1.25 $\pm$ 0.05) dex, respectively.  Note that both mean values are
distinguishable at 1-$\sigma$ level, i.e., the difference is $\Delta$($(V-I)_{o,-3}$) $\ge$ 0.10 mag, or
$\sigma$([Fe/H]) $\ge$ 0.3 dex.  We interprete such a metallicity 
bimodality  - seen along $\sim$ 5 Gyr -  as the possible evidence of the occurrence of a merger between
 two galaxies during their early epoch. One of them was more metal-poor and, due to the merger, its contributed gas was 
chemically enriched at a metallicity level similar 
to the other galaxy.  In addition, one of both galaxies
would have contributed to the resulting outer merged mass with much more gas and stars than the other, since the 
metal-rich peaks in the IC2 and OC Fornax fields contain twice up to four times more stars ($\sim$ more mass) 
than the metal-poor peak. In the innermost IC1 Fornax field, the contribution of the
supossed two colliding galaxies seems to be similar. 

Battaglia et al. \shortcite{betal06}, 
from spectroscopic metallicities and velocities of red giant stars at the central region of the galaxy,
also found evidence of a relatively recent merger of another galaxy or other means of gas accretion. 
Yozin \& Bekki \shortcite{yb12} using orbits for Fornax that are consistent with the latest proper motion 
measurements, showed that the observed asymmetric shell-like substructures can be formed 
from the remnant of a smaller dwarf during minor merging. 
%Recently, Lokas et al. \shortcite{letal14} have suggested an scenario of merger
%for And II, a dSph satellite of M31, with a resulting stable galaxy similar to Fornax. 
Nevertheless, since our results are constrained to three particular Fornax fields, extrapolation to other gaxaly 
regions requires further work. 

The merger of these two galaxies would have triggered a star formation bursting process that peaked between $\sim$ 6 
and 9 Gyr ago, depending on the position of the field in the galaxy.
We draw this conclusion from the comparison of the box sizes for each one of the AMRs along the
age axis. On the other hand, notice that the metallicity
spread after the burst is remarkable large, probably as a consequence of the presence
of not well-mixed gas out of which the new generation of star have been formed. During the most recent star formation
processes that we can account for (age $\sim$ 1-2 Gyr), the intrinsic metallicity dispersion results smaller than that
for the relatively older generations of stars.

We investigated the metallicity distributions for the three Fornax fields in three different galaxy lifetime 
periods, namely, 1-5 Gyr, 5-10 Gyr, and 10-15 Gyr, respectively. In order to obtain such distributions we 
associated each metallicity value in Fig. 4 to a gaussian distribution function with the mean [Fe/H] value, and
the errorbar and box sizes being the centre, the full width at half maximum, and the amplitude of the Gaussian,
respectively. We then summed the contribution of all the gaussian functions in the three lifetime periods. The results
are depicted in Fig. 5, where we distinguished those for the 1-5 Gyr, 5-10 Gyr, and 10-15 Gyr age  lifetime
periods with
solid, dotted and dashed lines, respectively. As can be seen, the bimodal metallicity distribution arises clearly
during the very earlier epoch of the galaxy evolution, while the most significant amount of stars have been
formed in the midst of its star formation history.

The AMRs recovered by del Pino et al. \shortcite{detal13} show a general good agreement with
the representative AMRs derived in this work (see Fig. 4), although they do not stand for
metallicity bimodality. The greatest difference occurs for the IC2 and OC fields at the age $\sim$ 9 Gyr. 
We recall that 
del Pino et al. considered all the observed stellar populations in the galaxy to recover the Fornax's 
AMR, whereas we only focused on those most numerous ones. This means that a representative population seen 
in a particular subfield could be blur whenever it is included in a composite stellar population framework,
so that its representative age/metallicity could result in a value in contrast with the global 
age/metallicity distribution. The meaning of the
"representative population" as a galaxy chemical evolution tracer, as well as how similar/different it is 
-in terms of galaxy chemical evolution- from that of a composite stellar population deserves further
studies which are beyond the scope of this paper; the representative metallicity bimodality shown in this work
being not altered.

\section{Summary}

From $VI$ photometric data in three fields of Fornax dSph obtained with $FORS1$ at the $VLT$, we produce
a comprehensive field star age-metallicity relationship (AMR) from the earliest epoch until $\sim$ 1 Gyr ago.
Although we find stars at any age in the three Fornax fields, the innermost one 
does not contains dominant very old stars (age $>$ 12 Gyr), whereas the relatively outer field does not account for representative
star field populations younger than $\sim$ 3 Gyr. Furthermore, when focusing on the most prominent stellar
populations, we find that the derived AMRs are engraved by the evidence of a outside-in star formation process 
in Fornax. 

On the metallicity side, the studied fields show bimodal distributions peaked at [Fe/H] = (-0.95 $\pm$ 0.15) dex 
and (-1.15 or -1.25 $\pm$ 0.05) dex, respectively, but only during the first half
of the entire galaxy lifetime. This is a possible evidence of the occurrence of a merger between two galaxies during 
their early epoch. In this context, one of both galaxies
would have contributed to the resulting outer merged mass with much more gas and stars than the other, while in the 
innermost Fornax field the contribution of the supossed two colliding galaxies seems to be similar.
We also find that the metallicity
spread $\sim$ 6 Gyr ago is remarkable large, probably as a consequence of the presence
of not well-mixed gas out of which the new generation of star have been formed. During the most recent star formation
processes that we can account for (age $\sim$ 1-2 Gyr), the intrinsic metallicity dispersion results smaller than that
for the relatively older generations of stars.  The merger of these two galaxies would have triggered a star 
formation bursting process that peaked between $\sim$ 6 
and 9 Gyr ago, depending on the position of the field in the galaxy.

\section*{Acknowledgements}
This work was partially supported by the Argentinian institutions CONICET and
Agencia Nacional de Promoci\'on Cient\'{\i}fica y Tecnol\'ogica (ANPCyT). 
We thank the anonymous referee, whose comments and suggestions have
allowed us to improve the manuscript.

\begin{figure}
\centerline{\psfig{figure=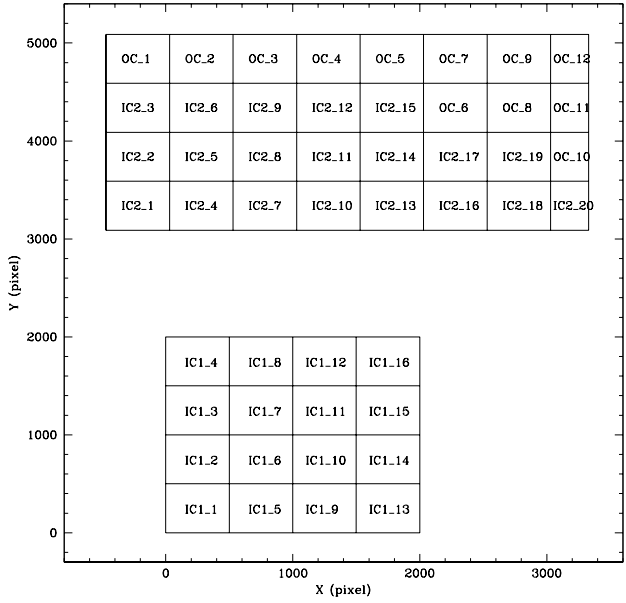,width=84mm}}
\caption{Schematic layout of the studied subfields.}
\label{fig1}
\end{figure}

\begin{figure}
\centerline{\psfig{figure=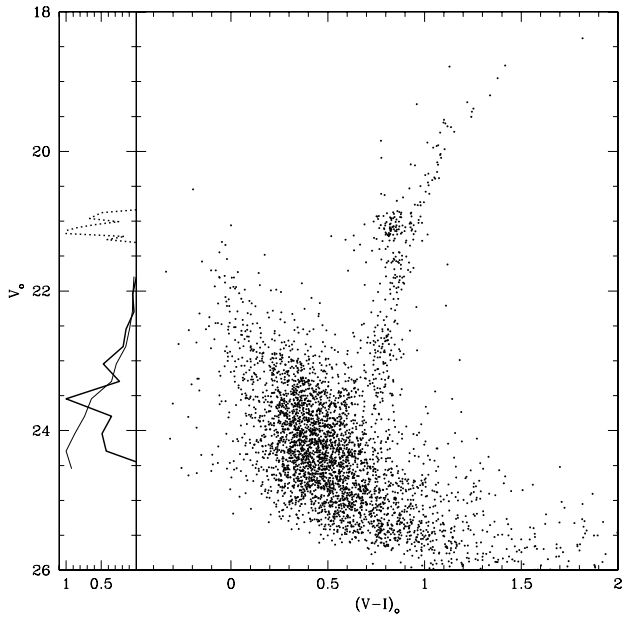,width=84mm}}
\caption{Hess diagram for the subfield IC1\_1. The normalized (thin solid line) and differential 
(thick solid line) 
MS luminosity functions as well as the RC luminosity distribution (dotted line) are shown.}
\label{fig2}
\end{figure}

\begin{figure}
\centerline{\psfig{figure=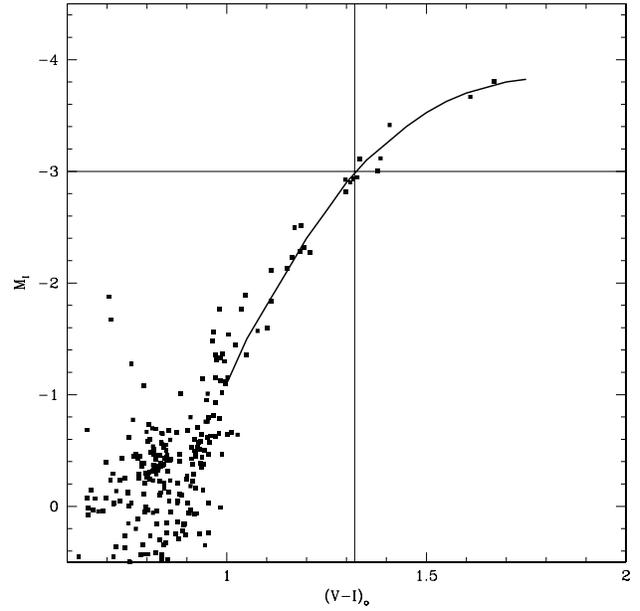,width=84mm}}
\caption{Enlargement of the Colour-Magnitude Diagram for stars in the IC1\_2 subfield with the
traced representative RGB sequence. The horizontal line corresponds to $M_I$ = -3.0 mag, and
the vertical line is placed at their intersection.}
\label{fig3}
\end{figure}

\begin{figure}
\centerline{\psfig{figure=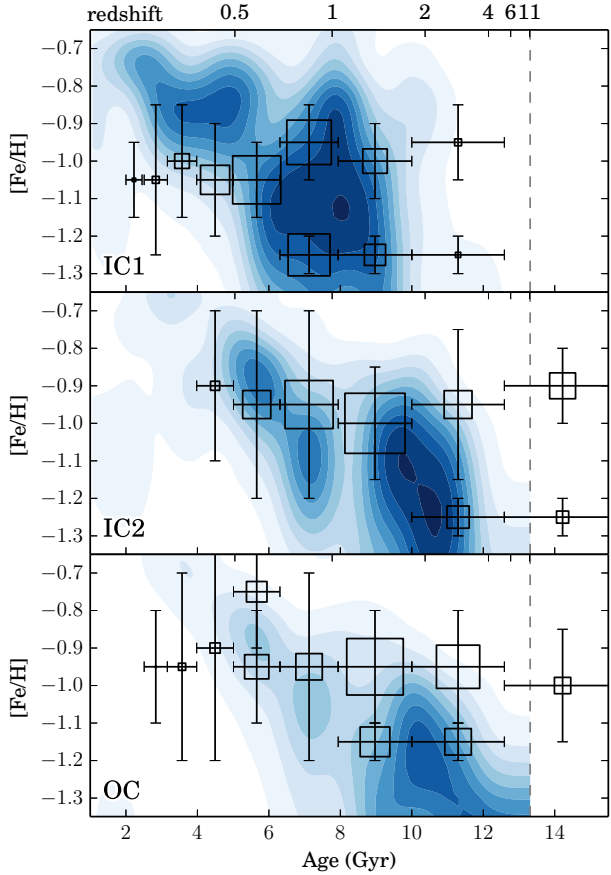,width=84mm}}
\caption{Mean Age-Metallicity relationships for the studied IC1, IC2, and OC Fornax fields.
The [Fe/H] errobars represent the intrinsic metallicity dispersion within the respective age 
intervals. The boxes drawn for each mean (age,[Fe/H]) point are proportional to the number of
subfields considered when averaging,  according to the procedure descripbed in Sect. 3, the age 
and metallicity values of Tables 3 and 4. The AMRs derived by del Pino et al. (2013) are also shown
as background Hess diagrams,
where the most numerous the stellar populations the bluer the contour colour.}
\label{fig4}
\end{figure}

\begin{figure}
\centerline{\psfig{figure=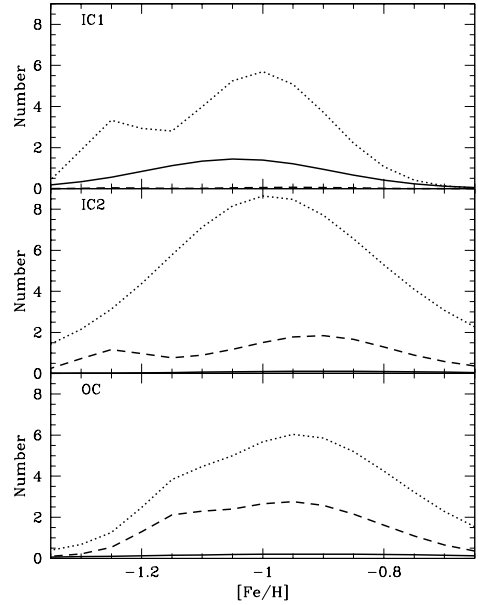,width=84mm}}
\caption{[Fe/H] distributions for three different galaxy lifetime periods: 1-5 Gyr (solid line), 5-10 Gyr (dotted line),
and 10-15 Gyr (dashed line).}
\label{fig5}
\end{figure}

\clearpage

\begin{table*}
\caption{Representative $V_o$(MSTO) magnitudes for the studied Fornax subfields.}
\begin{tabular}{@{}lcccccccccccccccccccc}\hline
Subfield  & 1 & 2 & 3 & 4 & 5 & 6 & 7 & 8 & 9 & 10  & 11 & 12 & 13 & 14 & 15 & 16 & 17 & 18 & 19 & 20\\
\hline
IC1 & 23.55 & 23.80 & 23.75 & 23.65 & 23.75 & 23.85 & 23.75 & 23.65 & 23.65 & 23.65 &  23.75 & 23.85 & 23.85 & 23.75 & 23.75 & 23.85 & --- & --- & --- &  ---\\
    & $\pm$0.20 & $\pm$0.30 & $\pm$0.50 & $\pm$0.40 & $\pm$0.30 & $\pm$0.30 & $\pm$0.40 & $\pm$0.50 & $\pm$0.20 & $\pm$0.30 & $\pm$0.30 & $\pm$0.40 & $\pm$0.20 & $\pm$0.30 & $\pm$0.30 & $\pm$0.30 & --- & --- & --- & ---\\
IC2 & 23.85 & 23.95 & 23.95 & 23.85 & 23.85 & 24.45 & 23.85 & 24.05  & 24.05 & 24.05 & 24.05 & 24.15 & 24.45 & 24.05 & 24.05 & 24.05 & 24.05 & 24.25 & 24.35 & 24.25 \\
    &$\pm$0.20 & $\pm$0.20 & $\pm$0.20 & $\pm$0.30 & $\pm$0.20 & $\pm$0.30 & $\pm$0.20 & $\pm$0.30 & $\pm$0.30 & $\pm$0.20& $\pm$0.30 & $\pm$0.30 & $\pm$0.20 & $\pm$0.30 & $\pm$0.30 & $\pm$0.20 & $\pm$0.30 & $\pm$0.30 & $\pm$0.20 & $\pm$0.20 \\
OC  & 24.25 & 24.05 & 24.25 & 23.85 & 24.25 & 23.85 & 24.25 & 24.05 & 24.05 & 24.35 & 24.25 & 24.25 & ---& ---& ---& ---& --- &--- &--- &---\\
    &$\pm$0.20 & $\pm$0.40& $\pm$0.40 &$\pm$0.30& $\pm$0.15 &$\pm$0.40& $\pm$0.30& $\pm$0.30& $\pm$0.30& $\pm$0.50  & $\pm$0.30& $\pm$0.15& --- &--- &--- &--- &--- &--- &--- &---\\\hline
\end{tabular}
\end{table*}

\begin{table*}
\caption{Representative $V_o$(RC) magnitudes for the studied Fornax subfields.}
\begin{tabular}{@{}lcccccccccccccccccccc}\hline
Subfield  & 1 & 2 & 3 & 4 & 5 & 6 & 7 & 8 & 9 & 10 & 11 & 12 & 13 & 14 & 15 & 16 & 17 & 18 & 19 & 20 \\
\hline
IC1 &  21.10 &21.10& 21.10& 21.10& 21.15& 21.10& 21.10& 21.10& 21.15& 21.10  & 21.10& 21.10& 21.15& 21.15& 21.10& 21.10& ---& ---& ---& ---\\
    & $\pm$0.05 &$\pm$0.05& $\pm$0.05& $\pm$0.05& $\pm$0.05& $\pm$0.05& $\pm$0.05& $\pm$0.05& $\pm$0.05& $\pm$0.05 & $\pm$0.05& $\pm$0.05& $\pm$0.05& $\pm$0.10& $\pm$0.05& $\pm$0.05& ---& ---& ---& ---\\ 
IC2 & 21.15 &21.15 &21.20& 21.20& 21.20& 21.25& 21.15& 21.20& 21.15& 21.15  & 21.20& 21.20& 21.15& 21.20& 21.20& 21.15& 21.15& 21.15& 21.15& 21.20\\ 
    & $\pm$0.05 &$\pm$0.05 &$\pm$0.05 &$\pm$0.05& $\pm$0.05& $\pm$0.05& $\pm$0.05& $\pm$0.05& $\pm$0.05& $\pm$0.10 & $\pm$0.10& $\pm$0.05& $\pm$0.10& $\pm$0.05& $\pm$0.10& $\pm$0.05& $\pm$0.05& $\pm$0.05& $\pm$0.05& $\pm$0.05\\
OC  &21.25 &21.20& 21.15& 21.25& 21.25& 21.25& 21.20& 21.20& 21.15& 21.20 & 21.20& 21.25& ---& ---& ---& ---& ---& ---& ---& ---\\
    & $\pm$0.05& $\pm$0.05& $\pm$0.05& $\pm$0.05& $\pm$0.05& $\pm$0.05& $\pm$0.05& $\pm$0.05& $\pm$0.05& $\pm$0.05 & $\pm$0.05& $\pm$0.05& ---& ---& ---& ---& ---& ---& ---& ---\\\hline

\end{tabular}
\end{table*}

\begin{table*}
\caption{Representative ages (Gyr) for the studied Fornax subfields.}
\begin{tabular}{@{}lcccccccccccccccccccc}\hline
Subfield  &  1&2&3&4&5&6&7&8&9&10 & 11&12&13&14&15&16&17&18&19&20\\
\hline
IC1&5.26&6.92&6.55&5.87&6.20&7.30&6.55&5.87&5.56&5.87 &6.55&7.30&6.92&6.20&6.55&7.30&---&---&---&---\\
&$\pm$1.45&$\pm$2.62&$\pm$3.93&$\pm$2.91&$\pm$2.38&$\pm$2.75&$\pm$3.21&$\pm$3.55&$\pm$1.53&$\pm$2.26 &$\pm$2.50&$\pm$3.54&$\pm$1.87&$\pm$2.72&$\pm$2.50&$\pm$2.75&---&---&---&---\\
IC2&6.92&7.70&7.30&6.55&6.55&11.64&6.92&8.12&8.56&8.56 &8.12&9.02&12.84&8.12&8.12&8.56&8.56&10.53&11.64&10.01\\
&$\pm$1.87&$\pm$2.06&$\pm$1.96&$\pm$2.50&$\pm$1.78&$\pm$4.04&$\pm$1.87&$\pm$3.01&$\pm$3.15&$\pm$2.70 &$\pm$3.44&$\pm$3.29&$\pm$3.74&$\pm$3.01&$\pm$3.44&$\pm$2.25&$\pm$3.15&$\pm$3.73&$\pm$2.89&$\pm$2.56\\
OC&9.50&8.12&10.53&6.20&9.50&6.20&10.01&8.12&8.56&11.07 &10.01&9.50&---&---&---&---&---&---&---&---\\
&$\pm$2.45&$\pm$3.88&$\pm$4.80&$\pm$2.38&$\pm$1.96&$\pm$3.06&$\pm$3.58&$\pm$3.01&$\pm$3.15&$\pm$6.11 &$\pm$3.58&$\pm$1.96&---&---&---&---&---&---&---&---\\
\hline
\end{tabular}
\end{table*}

\begin{table*}
\caption{Representative mean $(V-I)_o$ colours for the Red Giant Branch at $M_I$ = -3.0 mag for the studied Fornax subfields.}
\begin{tabular}{@{}lcccccccccccccccccccc}\hline
Subfield  & 1 & 2 & 3 & 4 & 5 & 6 & 7 & 8 & 9 & 10 & 11 & 12 & 13 & 14 & 15 & 16 & 17 & 18 & 19 & 20 \\
\hline
IC1 & 1.36 &1.32& 1.32& 1.38& 1.40& 1.38& 1.40& 1.38& 1.38& 1.38& 1.32& 1.40& 1.42& 1.40& 1.36& 1.32  & ---& ---& ---& ---\\ 
IC2 & 1.34 &1.44 &1.38 &1.44 &1.40 &1.32 &1.40& 1.38& 1.38& 1.42& 1.40& 1.40& 1.42& 1.40& 1.40& 1.40& 1.40& 1.40& 1.40& 1.40\\ 
OC  &1.40 &1.42& 1.40& 1.48& 1.40& 1.40& 1.40& 1.40& 1.40& 1.40& 1.36& 1.36& ---& ---& ---& ---& ---& ---& ---& ---\\\hline

\end{tabular}
\end{table*}

\begin{table*}
\caption{Representative [Fe/H] values for the studied Fornax subfields.}
\begin{tabular}{@{}lcccccccccccccccccccc}\hline
Subfield &  1&2&3&4&5&6&7&8&9&10 & 11&12&13&14&15&16&17&18&19&20\\
\hline
IC1&-1.10& -1.26&  -1.26 & -1.03 & -0.96 & -1.03 &-0.96 &-1.03 & -1.03& -1.03 &-1.26&-0.96 &-0.90&  -0.96&  -1.10&  -1.26&---&---&---&---\\ 
&  $\pm$0.07 &$\pm$0.08 &$\pm$0.08 &$\pm$0.07 &$\pm$0.07  &$\pm$0.07 & $\pm$0.07 &$\pm$0.07 &$\pm$0.07 &$\pm$0.07  &  $\pm$0.08&  $\pm$0.07& $\pm$0.06&  $\pm$0.07&  $\pm$0.07&  $\pm$0.08&---&---&---&---\\
IC2&-1.18 & -0.84& -1.03& -0.84 & -0.96 & -1.26& -0.96& -1.03 & -1.03 & -0.90  & -0.96&  -0.96& -0.90& -0.96&  -0.96&  -0.96&  -0.96&  -0.96&  -0.96&  -0.96\\ 
& $\pm$0.08 &$\pm$0.06& $\pm$0.07& $\pm$0.06& $\pm$0.07& $\pm$0.08 & $\pm$0.07 &$\pm$0.07 &$\pm$0.07 &$\pm$0.06 & $\pm$0.07 &  $\pm$0.07 & $\pm$0.06&  $\pm$0.07&  $\pm$0.07&  $\pm$0.07&  $\pm$0.07&  $\pm$0.07&  $\pm$0.07 &  $\pm$0.07\\
OC&-0.96 & -0.90&-0.96 &-0.73& -0.96 &-0.96& -0.96& -0.96 & -0.96 & -0.96 & -1.10&  -1.10&---&---&---&---&---&---&---&---\\ 
& $\pm$0.07 & $\pm$0.06 & $\pm$0.07 &$\pm$0.05 & $\pm$0.07 & $\pm$0.07 & $\pm$0.07 &$\pm$0.07  &$\pm$0.07  &$\pm$0.07 &  $\pm$0.07 &  $\pm$0.07&---&---&---&---&---&---&---&---\\
\hline
\end{tabular}
\end{table*}

%\begin{table*}
%\caption{Difference (in magnitude) between synthetic and observed (in that sense) representative $(V-I)_{o,-3}$ colours.}
%\begin{tabular}{@{}lcccccccccccccccccccc}\hline
%Subfield &  1&2&3&4&5&6&7&8&9&10 & 11&12&13&14&15&16&17&18&19&20\\
%\hline
%
%IC1 &  -0.06 &-0.02 &-0.04 &-0.08 &-0.12 &-0.06& -0.15& -0.10& -0.11& -0.08& -0.06& -0.14& -0.12& -0.12& -0.08& -0.04&---&---&---&\\
%IC2 &  -0.02 &-0.14 &-0.08 &-0.16 &-0.12 &-0.04& -0.10& -0.08& -0.10& -0.04& -0.12& -0.06& -0.14& -0.08& -0.14& -0.10& -0.12& -0.08& -0.12& -0.12\\
%OC  &  -0.08 &-0.14 &-0.10 &-0.15 &-0.15 &-0.15& -0.15& -0.12& -0.06& -0.08& -0.10& -0.04& ---  & ---&---&---&---&---&---&---\\
%\hline
%\end{tabular}
%\end{table*}
%
%\begin{table*}
%\caption{Difference (in dex) between synthetic and observed (in that sense) representative [Fe/H] values.}
%\begin{tabular}{@{}lcccccccccccccccccccc}\hline
%Subfield &  1&2&3&4&5&6&7&8&9&10 & 11&12&13&14&15&16&17&18&19&20\\
%\hline
%IC1 &-0.23& -0.08& -0.15& -0.30& -0.45& -0.23& -0.57& -0.38& -0.41& -0.30& -0.23& -0.53& -0.45& -0.45& -0.30& -0.15&   ---&---&---&--\\  
%IC2 &-0.08 &-0.53& -0.30 &-0.60& -0.45& -0.15& -0.38& -0.30& -0.38& -0.15 &-0.45& -0.23& -0.53& -0.30& -0.53& -0.38& -0.45& -0.30& -0.45\\ 
%OC &-0.30& -0.53& -0.38& -0.57& -0.57& -0.57& -0.57& -0.45& -0.23& -0.30& -0.38& -0.15& ---  & ---&---&---&---&---&---&---\\
%\hline
%\end{tabular}
%\end{table*}

\end{document}